\documentstyle[epsf]{mn}

\def\etal{{\rm et al.~}}
\def\eg{{\rm e.g.}}
\def\ie{{\rm i.e.}}

\title[Baryonic fraction in the CHDM universe]
{Baryonic fraction in the cold plus hot dark matter universe}

\author[E.~Choi and D.~Ryu]
{Eunwoo Choi$^{1}$\footnotemark[2] and Dongsu Ryu$^{2,3}$\footnotemark[2]\\ \\
$^1$Department of Astronomy \& Atmospheric Sciences,
    Kyungpook National University, Taegu 702-701, Korea\\
$^2$Department of Astronomy \& Space Science,
    Chungnam National University, Daejeon 305-764, Korea\\
$^3$Department of Astronomy, University of Washington,
    Box 351580, Seattle, WA 98185}

\date{Accepted 1997 ** **. Received 1997 ** **; in original form 1997 ** **}

\begin{document}

\maketitle

\begin{abstract}
We report a study to constrain the fraction of baryonic matter in the
cold plus hot dark matter (CHDM) universe by numerical simulations which
include the hydrodynamics of baryonic matter as well as the particle
dynamics of dark matter.
Spatially flat, {\it COBE}-normalized CHDM models with the fraction
of hot component $\Omega_h\leq0.2$ are considered.
We show that the models with $h/n/\Omega_h=0.5/0.9/0.1$ and $0.5/0.9/0.2$
give a linear power spectrum which agrees well with observations.
Here, $h$ is the Hubble constant in unit of $100~{\rm km/s/Mpc}$ and
$n$ is the spectral index of the initial power spectrum.
Then, for the models with $h/n/\Omega_h=0.5/0.9/0.2$ and baryonic fraction
$\Omega_b=0.05$ and $0.1$ we calculate the properties of X-ray clusters,
such as luminosity function, temperature distribution function,
luminosity-temperature relation, histogram of gas to total mass ratio,
and change of average temperature with redshift $z$.
Comparison with the observed data of X-ray clusters indicates that the
model with $\Omega_b=0.05$ is preferred.
The {\it COBE}-normalized CHDM model with $\Omega_b>0.1$ may be ruled
out by the present work, since it produces too many X-ray bright clusters.
\end{abstract}

\begin{keywords}
galaxies: clusters of -- hydrodynamics -- cosmology: large-scale structure
of Universe -- x-ray: general.
\end{keywords}

\footnotetext[2]{E-mail: ewchoi@vega.kyungpook.ac.kr;\\
ryu@sirius.chungnam.ac}

\section{Introduction}

The large-scale structure of the universe forms when matter accretes
onto high density perturbations via gravitational instability.
Among the models that attempt to explain the quantitative features of
the large-scale structure, the cold dark matter (CDM) model is based on
the assumption that the dominant component of matter is cold.
The model also assumes that the initial fluctuations are adiabatic and
random Gaussian, and have a Zel'dovich spectrum, $P(k) = Ak^n$ with
$n\sim1$.
The standard CDM (SCDM) model, which is the simplest and yet the most
widely studied, makes the additional assumptions that the total $\Omega = 1$
and $\Omega_b\sim0.05$ from the standard big bang nucleosynthesis
(Walker \etal 1991).
Here $\Omega$ is the density parameter. 
This model had some success, but it is now well known to have difficulties
in explaining a number of observations (see, \eg, Ostriker \& Steinhardt
1995).
In particular, this model has excessive power on small scales when
normalized to the {\it COBE} data on large scales (see, \eg, Bunn, Scott
\& White 1995). 
Several alternative cosmological models, including the CHDM model,
have been studied claiming better agreement with observations.

The CHDM model is based on the same assumptions as the SCDM model, except
that in addition to the CDM component, it has also the hot dark matter
(HDM) component.
HDM is composed of one of the three known species of neutrinos ($\nu_\tau$,
$\nu_\mu$, $\nu_e$) with a mass of $91.5\Omega_h h^2{\rm eV}$.
Since neutrino's free streaming erases fluctuations in density on
scales from galaxies to clusters during the radiation-dominated era,
replacing a portion of CDM with HDM suppresses power on small scales.
Therefore, the CHDM model has less power on small scales than the SCDM
model.
This basic property of the CHDM model was investigated some time ago
(Fang, Li \& Xiang 1984; Valdarnini \& Bonometto 1985; Achilli,
Occhionero \& Scaramella 1985).

The fact that the CHDM model is promising in explaining the observed
data of large-scale structure was first established by several analytic
calculations using linear or nonlinear tools (Schaefer, Shafi \& Stecker
1989; Van Dalen \& Schaefer 1992; Holtzman \& Primack 1993; Pogosyan \&
Starobinsky 1995; Dodelson, Gates \& Stebbins 1996; Ma 1996;
Borgani \etal 1995, 1997).
For example, Holtzman \& Primack (1993) used the peaks formalism for
Gaussian density fields and found that the correlation function of clusters
in the CHDM model with $\Omega_h=0.3$ is consistent with the correlation
function of the Abell clusters.

A number of numerical simulations further confirmed the promising
aspects of the CHDM model (Davis, Summers \& Schlegel 1992;
Klypin \etal 1993; Jing \etal 1994;  Klypin \& Rhee 1994; Nolthenius,
Klypin \& Primack 1994; Jing \& Fang 1994; Yepes \etal 1994; Bryan \etal
1994b; Ghigna \etal 1994; Bonometto \etal 1995; Walter \&  Klypin 1996;
Klypin, Nolthenius \& Primack 1997; Dav\'e \etal 1997).
Most of these simulations were carried out for the model with the
``standard'' density ratio $\Omega_c/\Omega_h/\Omega_b = 0.6/0.3/0.1$.
For example, Klypin \& Rhee (1994) calculated the correlation function
for the Abell clusters and the APM clusters and confirmed the results
given by Holtzman \& Primack (1993).
Jing \& Fang (1994) calculated the evolution of the mass function, the
velocity dispersion function, and the temperature function of clusters
and argued that their results favor the CHDM model over others.
Bryan \etal (1994b) computed the properties of X-ray clusters using
a code which can handle the hydrodynamics of baryonic matter.
They showed that the luminosity and temperature distribution functions
fit well the available observational data, and $\Omega_b\la0.1$ is
sufficient to explain the X-ray properties in the CHDM model with
$\Omega_h=0.3$.
Klypin \etal (1997) showed that the CHDM model with the ``standard'' 
density ratio, as well as that with $\Omega_h=0.2$, give good fits 
to a wide variety of ``present-epoch'' data.

However, the CHDM model with $\Omega_c/\Omega_h/\Omega_b = 0.6/0.3/0.1$
has a problem in producing galactic halos at $z\ga3$ massive enough to
account for neutral gas observed in damped Ly$\alpha$ systems
(Mo \& Miralda-Escud\'e 1994; Kauffmann \& Charlot 1994; Ma \& Bertschinger
1994; Klypin \etal 1995; Bi, Ge \& Fang 1995).
So it was suggested that the more promising version of the CHDM model should
have less mass in the hot component, $\Omega_h\la0.2$, in order to
have more power on galaxy scales and so produce significantly more
high-redshift objects.
Ma (1996), using an approximation to the evolution of
linear power spectrum, argued that the slightly tilted ($n\sim0.9-0.95$)
CHDM model with $\Omega_h\sim0.1-0.2$ gives a power spectrum that agrees
best with the observed power spectrum.
Primack \etal (1995) showed that dividing the neutrino mass between two
species of neutrinos ($\nu_\tau$, $\nu_\mu$) lowers the power spectrum 
on the cluster scale, and thus lowers the cluster abundance without 
the necessity of tilt $n < 1$ of the CHDM model.

In this paper, we constrain $\Omega_b$ in the CHDM models with
$\Omega_h\leq0.2$ by investigating the properties of X-ray clusters
that are sensitive to $\Omega_b$.
Recent satellite X-ray observations made the properties of X-ray clusters
increasingly important to cosmology as a probe into the large-scale
structure of the universe.
Being massive and rare, the cluster abundance in the local and distant
universe carries vital information on the initial density fluctuations
and the matter content of the universe.
Also being relatively young dynamically, the details of their structures
provide us with some signatures left over from the formation epoch as well
as information on the background cosmology.
The study on the properties of X-ray clusters in model universe has been
made analytically (\eg, Kitayama \& Suto 1996) or numerically using
grid-based codes (Kang \etal 1994; Cen \& Ostriker 1994; Bryan \etal 1994a;
Bryan \etal 1994b) and the SPH code (\eg, Evrard, Metzler \& Navarro 1996).
A general consensus seems to be that the SCDM model with $h=0.5$ normalized
to the {\it COBE} measurement of the anisotropies in the cosmic background
radiation (\ie, $\sigma_8>1$) has serious difficulties in explaining
the observed properties of X-ray clusters, such as the cluster abundance
(Kang \etal 1994; Bryan \etal 1994a), the baryon fraction in clusters
(Lubin \etal 1995), and the contribution of cluster emission to the X-ray
background (Kang \etal 1994; Bryan \etal 1994a; Kitayama \& Suto 1996).
On the other hand, a flat CDM model with cosmological constant ($\Lambda$CDM)
and a CHDM model, both of which have smaller values of $\sigma_8$, seem more
consistent with observations (Cen \& Ostriker 1994; Bryan \etal 1994b).
Here, $\sigma_8$ is the present-epoch, linear rms mass fluctuation
in the sphere of radius $8 h^{-1}{\rm Mpc}$.

We should mention that our work bears many similarities with
that of Bryan \etal (1994b).
Both consider X-ray clusters in the CHDM universe with similar (but
not same) codes.
However, while Bryan \etal (1994b) considered the CHDM model with
$\Omega_c/\Omega_h/\Omega_b = 0.6/0.3/0.1$, which was ``standard''
at that time, we considered the models with a smaller $\Omega_h$
($\Omega_h=0.2$, see \S 3.2), which are more acceptable now.

In \S 2 the details of the numerical simulations are described.
Results are presented in \S 3.
Conclusion follows in \S 4.

\section{Numerical Simulations}

\subsection{Choice of Models}

In order to study the linear power spectrum using analytic approximations
in the CHDM universe, Ma (1996) considered the models with
$0.5 \leq h \leq 0.8$, $0.8 \leq n \leq 1$, and
$0.05 \leq \Omega_h \leq 0.3$.
The normalization of the power spectrum, $\sigma_8$, was
determined by the {\it COBE} observation (see \S 2.2 for details).
As mentioned in Introduction, Ma (1996) found that the models with
$n \sim 0.9-0.95$ and $\Omega_h \sim 0.1-0.2$ give power spectra
that agree well with observation.
Here, we adopt Ma's models with $h/n/\Omega_h=0.5/0.9/0.1$ and
$0.5/0.9/0.2$.
For baryonic fraction we use the values $\Omega_b=0.05$ and $0.1$,
which is in the range predicted by big bang nucleosynthesis
$0.007h^{-2} \la \Omega_b \la 0.024h^{-2}$ (Walker \etal 1991;
Copi, Schramm \& Turner 1995).
The values of the model parameters are summarized in Table 1.

\begin{table}
\caption{Model Parameters.}
\begin{tabular}{ccccc}
 Parameter  & CHDMa   & CHDMb   & CHDMc   & CHDMd   \\
\hline 
 $h$        & 0.5     & 0.5     & 0.5     & 0.5     \\ 
 $n$        & 0.9     & 0.9     & 0.9     & 0.9     \\ 
 $\Omega_c$ & 0.75    & 0.7     & 0.85    & 0.8     \\ 
 $\Omega_h$ & 0.2     & 0.2     & 0.1     & 0.1     \\ 
 $\Omega_b$ & 0.05    & 0.1     & 0.05    & 0.1     \\ 
 $\sigma_8$ & 0.67688 & 0.67688 & 0.73921 & 0.73921 \\
\end{tabular}
\end{table}

\subsection{Initial Power Spectrum}

Assuming the density distribution is Gaussian random, the initial condition
is determined by the power spectrum only.
It is given by the following functional form
\begin{equation}
P(k,a,\Omega_h) = a^2Ak^nT^2(k,a,\Omega_h)_,
\end{equation}
where $a = (1+z)^{-1}$ is the expansion parameter, $A$ is a normalization 
constant, and $k$ is the wave number.
$T(k,a,\Omega_h)$ is the transfer function describing linear changes
in the perturbation.
We adopt the transfer function derived by Ma (1996).
For CDM, it is given by
\begin{displaymath}
T_c(k,a,\Omega_h) = \frac{\ln(1+2.34q)}{2.34q}
\end{displaymath}
\begin{displaymath}
\times \left[\frac{1}{1+3.89q+(16.1q)^2+(5.46q)^3+(6.71q)^4}\right]^{1/4}
\end{displaymath}
\begin{equation}
\times \left[\left(\frac{1+0.01647x^{3.259/2}+2.803 \times 10^{-5}
x^{3.259}} {1+10.90x_0^{3.259}}\right)^{\Omega_h^{1.05}}\right]^{1/2}_,
\end{equation}
where $\Gamma = \exp(-2\Omega_b)h$, $q = k/\Gamma h$, $\Gamma_\nu =
a^{1/2}\Omega_h h^2$, $x = k/\Gamma_\nu$, $x_0 = x(a=1)$, and $k$ is
in unit of ${\rm Mpc}^{-1}$.
For HDM, it is given by
\begin{equation}
T_h(k,a,\Omega_h) = T_c(k,a,\Omega_h)
\end{equation}
\begin{displaymath}
\times \left[\frac{\exp(-0.0015x')}{1-0.121x'^{1/2}
+0.102x'-0.0162x'^{3/2}+0.00171x'^2}\right]^{1/2}_,
\end{displaymath}
where $x' = k / \Gamma_\nu h$.
We assume that CDM and baryonic matter have the same power spectrum.
Then, the density-weighted power spectrum
$P = [\Omega_h P_h^{1/2} +(1-\Omega_h)P_c^{1/2}]^2$ describes the
gravitational perturbations contributed by all the matter components.

The normalization constant $A$ is related to $\sigma_8\equiv\sigma
(R=8h^{-1}{\rm Mpc},a=1,\Omega_h)$ by the following relation
\begin{equation}
\sigma^2(R,a,\Omega_h) = {\int_0^\infty \frac{dk}{k} 4\pi k^3 P(k,a,\Omega_h)
W^2(kR)}_,
\end{equation}
where $W(kR) = 3[\sin(kR) - kR\cos(kR)]/(kR)^3$ is the top-hat window
function.
After {\it COBE}, the rms quadrupole $Q_{rms-PS}$ inferred from the cosmic
microwave background anisotropy is often used to fix $\sigma_8$.
In the CHDM universe, $\sigma_8$ is approximately given by
\begin{displaymath}
\sigma_8 = Q_{18}(n)~\frac{h^2 0.008^{(1-n)/2}}{0.0136+0.294h^{0.803}
+0.109h^2}
\end{displaymath}
\begin{equation}
\hspace{22pt} \times~\frac{1+(1-\Omega_h)^{5.11} 0.116(\Omega_h h)^{-0.893}}
{1+0.116(\Omega_h h)^{-0.893}}_,
\end{equation}
where $Q_{18}(n) = Q_{rms-PS}(n)/18\mu{\rm K}$ (Ma 1996).
We take $Q_{rms-PS}=18\mu{\rm K}$ for $n = 1$, $Q_{rms-PS}=19.2\mu{\rm K}$
for $n = 0.9$, and $Q_{rms-PS}=20.5\mu{\rm K}$ for $n = 0.8$, respectively
(G\'orski \etal 1994; Ma 1996).

\subsection{Numerical Method}

The simulations have been performed with the cosmological hydrodynamic
code described in Ryu \etal (1993).
It is based on the Total Variation Diminishing (TVD) scheme, which is an
explicit, second-order, Eulerian finite difference scheme (Harten 1983).
Extra care was taken with the code in two aspects:
(1) Strong shocks (with Mach number larger than $\sim100$) were identified
and handled specially in order to prevent the pre-shock regions from being
heated unphysically.
(2) Self-gravity and cosmological expansion were included in the way that
the total energy conservation represented by the Layzer-Irvine equation
(Peebles 1980) is preserved.

Since the core radius of typical X-ray clusters is $\la0.5h^{-1}{\rm Mpc}$
and the cluster-cluster separation is $\sim50h^{-1}{\rm Mpc}$, $\sim100^3$
is the absolute minimum number of grids required for baryonic matter in
order to get statistically meaningful X-ray quantities. 
The simulations have been done in a periodic computational domain using
$256^3$ grids, $128^3$ CDM particles, and $2 \times 128^3$ HDM particles.
The initial condition for the HDM particles has been generated in pairs
with random and opposite thermal velocities in order to sample neutrino
phase space distribution (Klypin \etal 1993).
The comoving size of the computational domain is $80h^{-1} {\rm Mpc}$, so
the grid size is $0.31h^{-1}{\rm Mpc}$.
While a smaller size would allow us to resolve cluster structure better,
a larger size would allow larger waves included and give us a larger sample
of high-luminosity and high-temperature clusters. 
With the domain size of $80h^{-1}{\rm Mpc}$, which has been determined by a
compromise between the two considerations, clusters are identified
comfortably and yet enough clusters are yielded for statistical analyses.
However, with the grid size of $0.31h^{-1}{\rm Mpc}$, clusters are not
``fully'' resolved resulting in under-estimate of X-ray luminosities (see
\S 3.2 for further discussion).
The simulations have been started at $z = 15$ (or $a = 1/16$) and run
to $z = 0$ (or $a = 1$).

In the simulations, atomic processes such as heating and cooling have not
been included, as they will have little effect on the hot cluster gas
discussed in this paper.

\section{Results}

\subsection{Power Spectrum}

We have calculated the linear power spectrum of total (dark and baryonic)
matter for the 4 models listed in Table 1 using the formulae in \S 2.2.
Figure 1 shows them as well as the power spectrum reconstructed from 
galaxy and cluster surveys by Peacock \& Dodds (1994).
The upper panel contains the power spectra of the models with
$h/n/\Omega_h=0.5/0.9/0.1$ (Model CHDMc and CHDMd) and the lower panel with
$h/n/\Omega_h=0.5/0.9/0.2$ (Model CHDMa and CHDMb).
They agree well with the Peacock \& Dodds's reconstructed power spectrum as
was already shown by Ma (1996), although the Peacock \& Dodds's power
spectrum is for the CDM model.
Note that each panel contains a dotted curve for the model with
$\Omega_b=0.05$ and a solid curve for the model with $\Omega_b=0.1$,
although they are not well distinguished.
In general, increasing $\Omega_b$ with fixed $\Omega_h$ results in decreasing
power at small scales, but only by a small amount.
The plots indicate that the power spectrum is insensitive to the baryonic
fraction.
The power spectra of other CHDM models with $0.5 \leq h \leq 0.8$,
$0.8 \leq n \leq 1$, and $0.05 \leq \Omega_h \leq 0.3$ can be found in
Choi (1996).

\begin{figure*}
\vspace{1.25truein}
\epsfysize=4in\epsfbox[50 300 500 600]{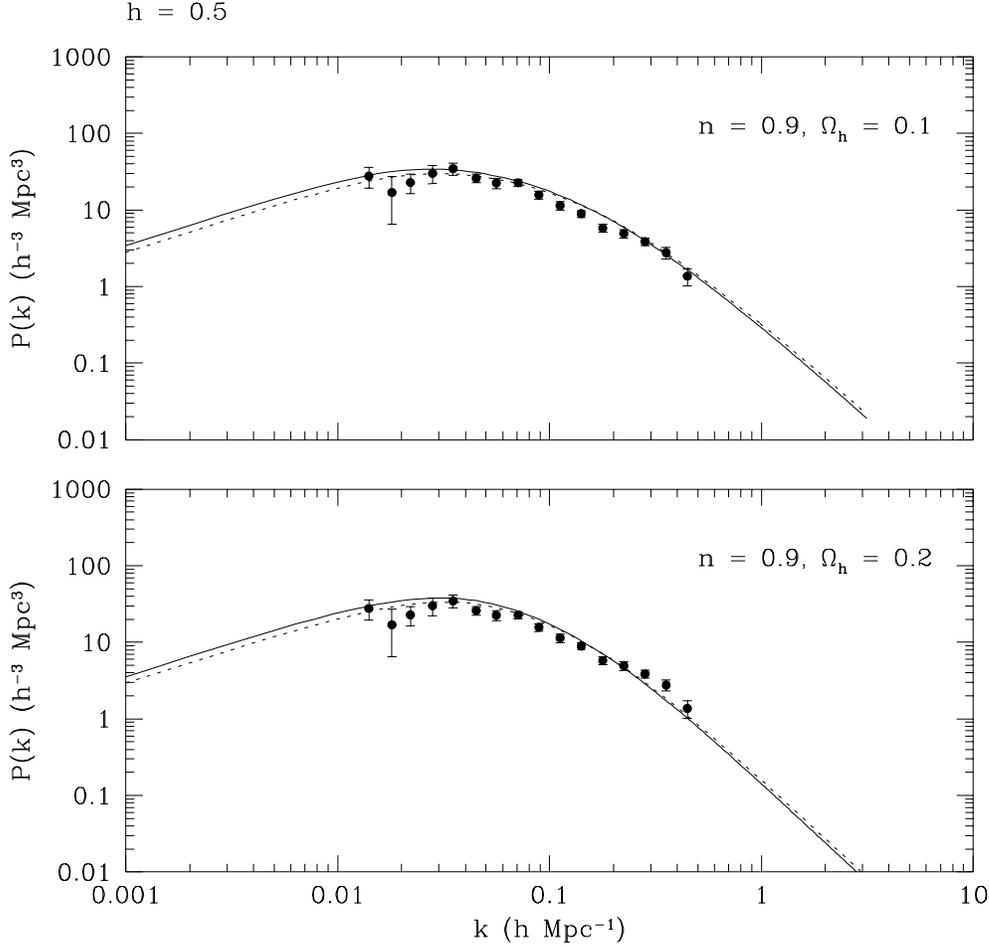}
\vspace{-0.3truein}
\caption{Linear power spectra of total (dark and baryonic) matter for
Model CHDMc and CHDMd (upper panel) and CHDMa and CHDMb (lower panel).
Each panel contains a dotted curve for the model with $\Omega_b=0.05$
(Model CHDMc and CHDMa) and a solid curve for the model $\Omega_b=0.1$
(Model CHDMd and CHDMb).
The reconstructed power spectrum from galaxy and cluster surveys is shown
with filled circles (Peacock \& Dodds 1994).}
\end{figure*}

\subsection{Properties of X-ray Clusters}

The properties of X-ray clusters are sensitive to $\Omega_b$, since the
X-ray bremsstrahlung emission is proportional to the square of gas density.
So they can be used to constrain $\Omega_b$.
We select two models, Model CHDMa and CHDMb with $h/n/\Omega_h=0.5/0.9/0.2$
and $\Omega_b=0.05$ and $0.1$, and study by numerical simulations the
properties of X-ray clusters in those models in detail.

X-ray clusters emit X-rays from the hot intra-cluster gas that fills the
space between galaxies and occupies much of the cluster's volume.
In general the core radius of typical X-ray clusters is about 
$0.5h^{-1}{\rm Mpc}$ and the total radius about $1h^{-1}{\rm Mpc}$,
but they cannot be accurately determined from observations. 
The total (bolometric) X-ray luminosity due to the thermal bremsstrahlung is
\begin{equation}
L_x = {4 \pi \int_\nu \int_V j_{ff} ~ d^3{\bf r} ~ d\nu}_,
\end{equation}
where $j_{ff}$ is given in units of ${\rm erg}~{\rm cm}^{-3}
{\rm s}^{-1} {\rm Hz}^{-1} {\rm sr}^{-1}$ by
\begin{displaymath}
j_{ff} = \frac{1}{4 \pi}~\frac{32e^4}{3m_e^2c^3}\left[\frac{\pi h
\nu_0({\rm H})}{3kT}\right]^{1/2}\exp\left(-\frac{h\nu}{kT}\right)
\end{displaymath}
\begin{equation}
\hspace{27pt} \times~g_{ff}(T,\nu)\left[n({\rm HII})+n({\rm HeII})+4n({\rm HeIII})
\right]n(e).
\end{equation}
$g_{ff}(T,\nu)$ is the Gaunt factor.
We assume the primordial abundance, $76\%$ H and $24\%$ He by mass, and that
H and He are fully ionized.

X-ray clusters in the simulations are identified as follows.
We first calculate the total X-ray luminosity due to the thermal
bremsstrahlung in each cell.
The cells with the total X-ray luminosity higher than
$10^{38}{\rm erg}~{\rm s}^{-1}$ are selected as X-ray bright cells.
Then, we find the local maxima by comparing the total X-ray luminosity
of each X-ray bright cell with that of $124$ neighboring cells, and
identify them as the centers of the X-ray clusters. 
Having identified the X-ray cluster centers, we define the X-ray 
clusters in the whole computational domain.
The total X-ray cluster volume  consists of $125$ cells, the central cell
and $124$ cells surrounding it.
For the calculations of X-ray luminosity and temperature, each cell
is weighted equally by a weight factor $ = 4\pi/3 ~ $
(radius)$^3$/(resolution)$^3$/(the number of cells)
$ = 4\pi/3 ~ (1.0)^3/(0.31)^3/125 = 1.098$, so that the total volume
of each X-ray cluster equals the volume of a sphere of radius
$1h^{-1} {\rm Mpc}$.
The weighting scheme compensates for the adoption of a slightly small
volume by heightening the weight per cell.

\begin{figure*}
\caption{Projected distribution of the X-ray clusters with
$L_x >10^{41} {\rm erg}~{\rm s}^{-1}$ from the region of
$r\le1h^{-1}{\rm Mpc}$.
The open circles in the left panels represent the X-ray clusters in CHDMa,
and the filled circles in the right panels represent the X-ray clusters
in CHDMb.}
\end{figure*}

\begin{figure*}
\vspace{0.4truein}
\epsfysize=4.2in\epsfbox[-30 200 370 600]{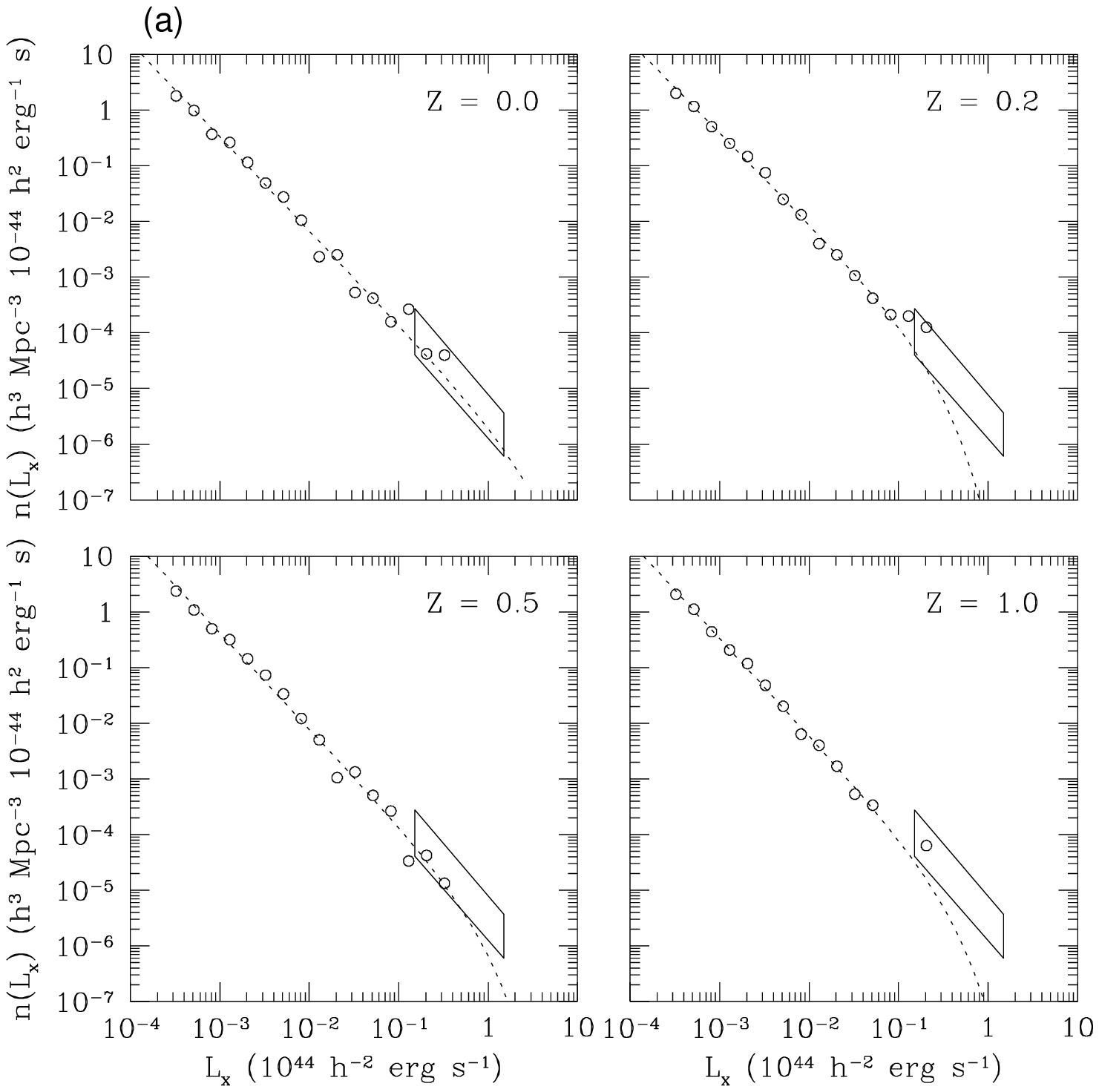}
\vspace{0.3truein}
\epsfysize=4.2in\epsfbox[-30 200 370 600]{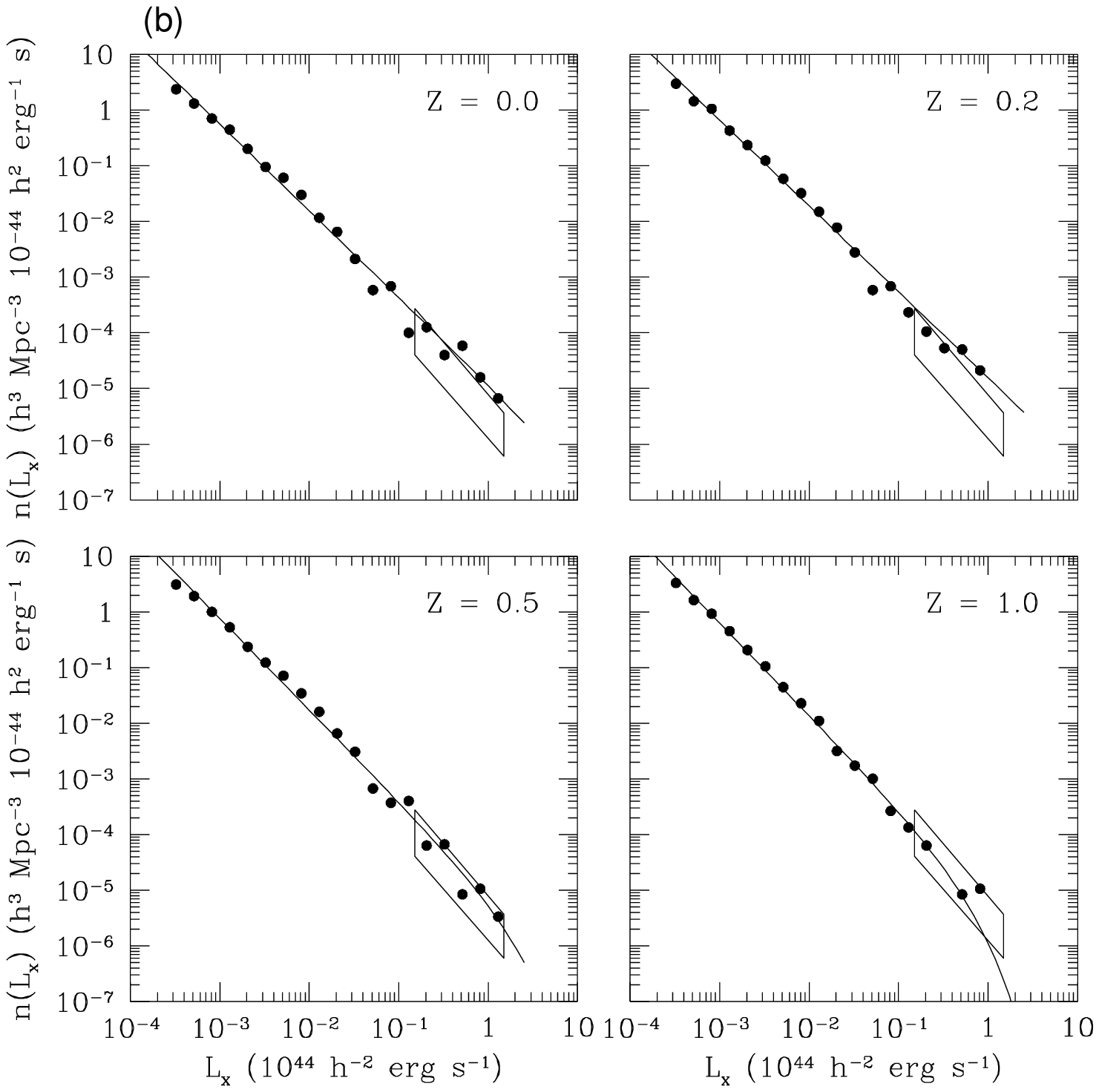}
\vspace{-0.3truein}
\caption{Luminosity function of the X-ray clusters at four
different epochs with X-ray luminosity from the region of
$r\le1h^{-1}{\rm Mpc}$ integrated over the whole frequency range. 
The open circles in (a) represent the luminosity function in CHDMa,
and the filled circles in (b) represent the luminosity function in CHDMb.
The dotted and solid lines are the Schechter fits, and the boxed area shows
the observational data (Henry \& Arnaud 1991).}
\end{figure*}

We note that our scheme is similar to (but not same as) that used by
Kang \etal (1994), Cen \& Ostriker (1994), and Bryan \etal (1994a),
but different from that used by Bryan \etal (1994b).
Our test showed that the calculated luminosity and temperature are
insensitive to the details of the scheme, as already pointed by
Bryan \etal (1994b).
However, as pointed in those previous works, our luminosity should have
been under-estimated due to insufficient resolution to resolve the
central density structure in clusters.
Bryan \etal (1994b) pointed that in their simulation with resolution similar
to ours, the under-estimation in the luminosity could be as high as
a factor of $3$.
The importance of resolution was also point by Evrard, Metzler \& Navarro
(1996) in the study of X-ray clusters using the SPH code.
On the other hand, the calculated temperature is relatively insensitive to
resolution.

The bright X-ray clusters with $L_x > 10^{41} {\rm erg}~{\rm s}^{-1}$ from
the region of radius $1h^{-1}{\rm Mpc}$ at $z = 0$ are shown in Figure 2.
Left panels show the projection of the X-ray clusters generated in CHDMa
in the $x-y$, $y-z$, and $z-x$ planes, and the right panels show
the projection of the X-ray clusters generated in CHDMb.
The distribution pattern of the X-ray clusters is similar, but there are more
X-ray clusters in CHDMb (with larger $\Omega_b$) than in CHDMa. 
In CHDMa, our computational domain with $(80h^{-1}{\rm Mpc})^3$ volume
at $z = 0$ contains no cluster with total luminosity brighter than
$10^{45} {\rm erg}~{\rm s}^{-1}$, no cluster brighter than
$10^{44} {\rm erg}~{\rm s}^{-1}$, $13$ clusters brighter than
$10^{43} {\rm erg}~{\rm s}^{-1}$, $44$ clusters brighter than
$10^{42} {\rm erg}~{\rm s}^{-1}$, and $268$ clusters brighter than
$10^{41} {\rm erg}~{\rm s}^{-1}$.
In CHDMb, our computational domain at $z = 0$ contains no cluster with
total luminosity brighter than $10^{45} {\rm erg}~{\rm s}^{-1}$,
$2$ clusters brighter than $10^{44} {\rm erg}~{\rm s}^{-1}$,
$24$ clusters brighter than $10^{43} {\rm erg}~{\rm s}^{-1}$,
$126$ clusters brighter than $10^{42} {\rm erg}~{\rm s}^{-1}$, and
$558$ clusters brighter than $10^{41} {\rm erg}~{\rm s}^{-1}$.

The luminosity functions calculated with the identified X-ray clusters are
shown in Figures 3a and 3b for CHDMa and CHDMb. 
The plots show the luminosity functions in
$10^{40}{\rm erg}~{\rm s}^{-1} \leq L_x \leq 10^{45} {\rm erg}~{\rm s}^{-1}$
and $0 \leq z \leq 1$.
There are very few high-luminosity clusters with
$L_x \ga 10^{44} {\rm erg}~{\rm s}^{-1}$, because our computational domain
is too small to contain them.
Also the plots show the observed bolometric luminosity function
$\{3.1_{-1.8}^{+4.5} \times 10^{-6} h^3 {\rm Mpc}^{-3}
h^2 [L_{44}(bol)]^{-1}\} [h^2 L_{44}(bol)]^{-1.85 \pm 0.4}$ by Henry \&
Arnaud (1991) as the boxed area.
The Henry \& Arnaud's luminosity function agrees with the more recently
considered ones, such as the one by Burn \etal (1996), although
White, Efstathiou \& Frenk (1993) pointed errors in the Henry \& Arnaud's
luminosity function.
The observed bolometric luminosity function is more consistent with 
the calculated luminosity function of CHDMa than that of CHDMb.
A possible correction for the under-estimation of the X-ray luminosity
would make the discrepancy between the calculated luminosity function
of CHDMb and the observed luminosity function worse.

Also in CHDMb we find weak positive evolution until $z \sim 0.2$, and
then mild negative evolution thereafter.
This can be seen in Figure 4 which shows the number density evolution of
X-ray clusters with $L_x > 10^{43} {\rm erg}~{\rm s}^{-1}$ between
$z = 0$ and $z = 1.5$.
Open circles are for CHDMa and filled circles are for CHDMb.
We note that the evolution of the cluster abundance does not agree
with the analytic predictions such as the one based on the Press-Schechter
approximation (Bahcall, Fan \& Cen 1997; Fan, Bahcall, \& Cen 1997),
according to which the evolution is much steeper in high $\Omega$ and
low $\sigma_8$ models.

\begin{figure}
\vspace{0.5truein}
\epsfysize=2.4in\epsfbox[160 300 460 500]{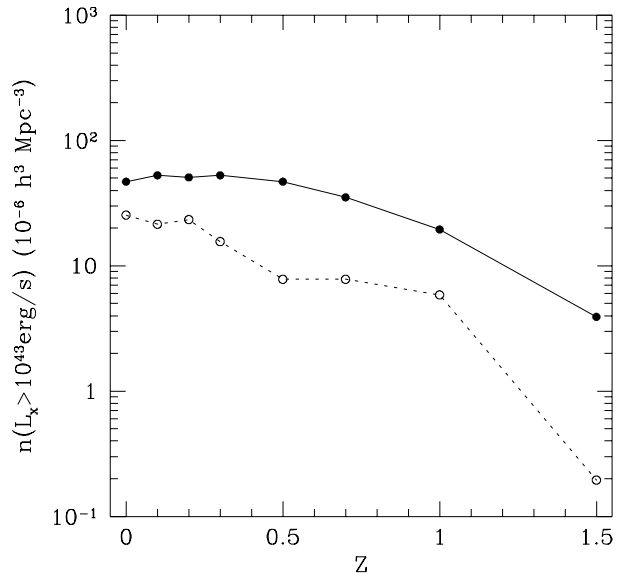}
\caption{Evolution of the number density of the X-ray clusters with
$L_x > 10^{43} {\rm erg}~{\rm s}^{-1}$ from the region of
$r\le1h^{-1}{\rm Mpc}$.
The open circles and dotted line are for CHDMa, and the filled circles and
solid line are for CHDMb.}
\end{figure}

We fit the luminosity function to an approximate Schechter function
\begin{equation}
n(L)dL \equiv n_0 \left(\frac{L}{L^*}\right)^{-\alpha} \exp\left(
-\frac{L}{L^*}\right) \frac{dL}{L^*}
\end{equation}
using the Levenberg-Marquardt method, and determine the Schechter 
parameters ($\alpha$, $L^*$, $n_0$) as a function of redshift. 
The parameters are calculated at seven different redshifts and 
listed in Table 2.
Here, $L^*$ is in unit of $10^{44} {\rm erg}~{\rm s}^{-1}$ and
$n_0$ is in unit of $10^{-6}h^3{\rm Mpc}^{-3}$.
The Schechter functions with these parameters are plotted with dotted
lines for CHDMa and solid lines for CHDMb in Figure 3. 
The value of the Schechter $\alpha$-parameter in CHDMa, $\sim1.7$,
is slightly larger than that in CHDMb, but still smaller than the
best-fit observational values, $1.9 - 2.0$ quoted by Henry (1992). 
However, the Schechter $\alpha$-parameters in our simulations are
primarily determined by lower luminosity clusters than those used for
the fits of observational data (Henry 1992).
The characteristic luminosity, $L^*$, and the total number density,
$n_0$, of the Schechter luminosity function are determined less reliably,
because the number of samples is too small.

\begin{table}
\caption{Schechter fits for the X-ray clusters luminosity function.}
\begin{tabular}{cccccc}
 Model & $~~~~~~z~~~~~~$ & $~~~~~~\alpha~~~~~$ & $~~~~~L^*~~~~$ &
$~~~~~n_0~~~~$ \\
\hline 
       & 0.0 & 1.69 &  2.51 &  1.48 \\
       & 0.1 & 1.55 &  0.04 & 47.27 \\
       & 0.2 & 1.64 &  0.19 & 13.76 \\
 CHDMa & 0.3 & 1.61 &  0.30 & 11.57 \\
       & 0.5 & 1.72 &  0.67 &  3.85 \\
       & 0.7 & 1.71 &  0.31 &  6.64 \\
       & 1.0 & 1.74 &  0.29 &  4.95 \\
\hline 
       & 0.0 & 1.56 & 14.48 &  2.69 \\
       & 0.1 & 1.56 & 23.22 &  2.37 \\
       & 0.2 & 1.54 & 86.69 &  1.41 \\
 CHDMb & 0.3 & 1.57 &  1.09 & 12.96 \\
       & 0.5 & 1.65 &  1.87 &  5.89 \\
       & 0.7 & 1.67 &  2.03 &  4.60 \\
       & 1.0 & 1.67 &  0.57 &  9.31 \\
\end{tabular}
\end{table}

\begin{figure*}
\vspace{0.4truein}
\epsfysize=4.2in\epsfbox[-30 200 370 600]{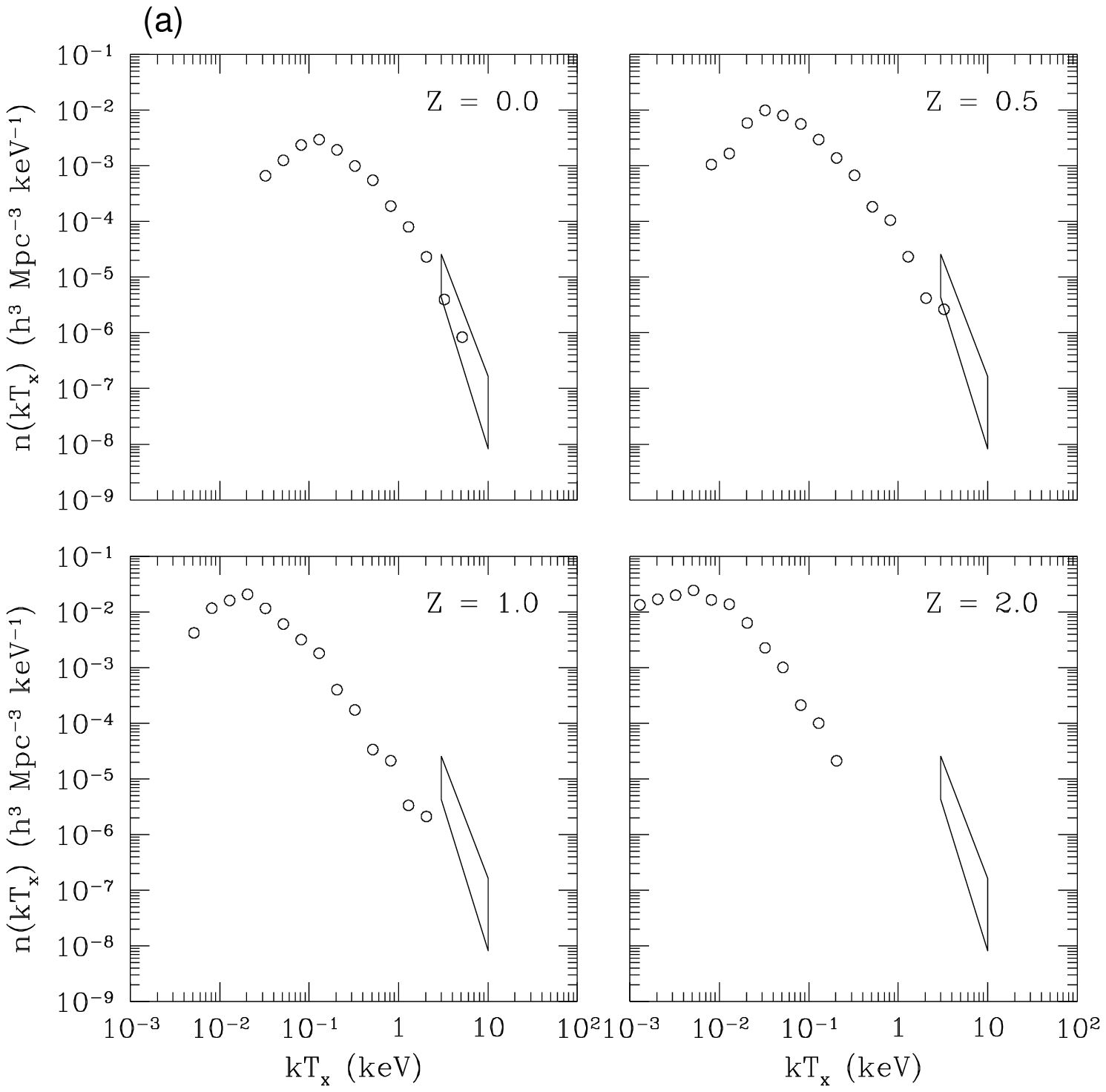}
\vspace{0.3truein}
\epsfysize=4.2in\epsfbox[-30 200 370 600]{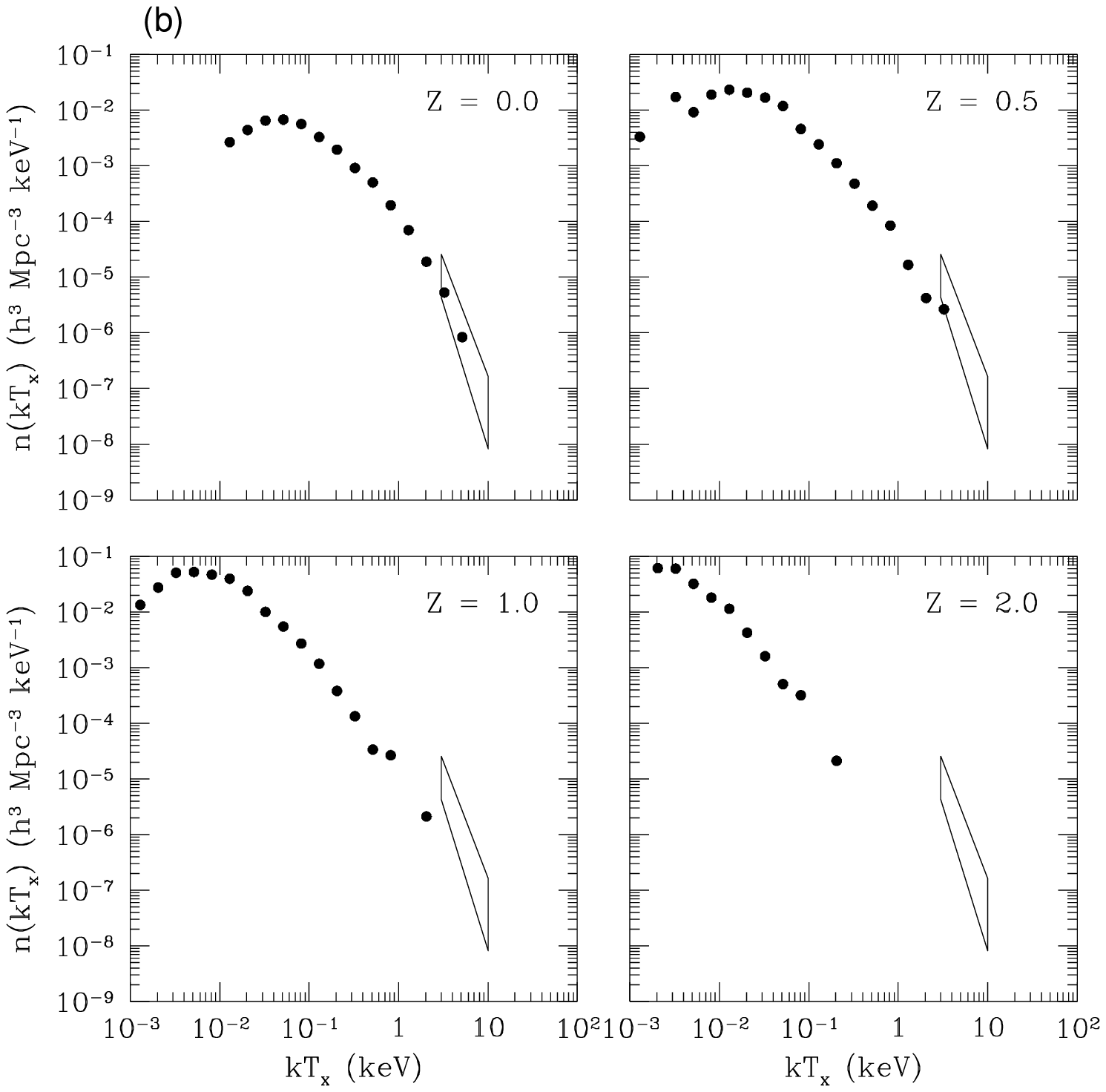}
\vspace{-0.3truein}
\caption{Temperature distribution function of the X-ray clusters
at four different epochs.
The open circles in (a) represent the distribution function in CHDMa,
and the filled circles in (b) represent the distribution function in CHDMb.
The boxed area shows the observation data (Henry \& Arnaud 1991).}
\end{figure*}

The temperature distribution function, in addition to the luminosity
function, of X-ray clusters provides a useful test for the structure
formation theory. 
It has been calculated at four different redshifts and shown in Figures
5a and 5b for CHDMa and CHDMb. 
The turnover at low X-ray temperature, $T_x$, is possibly caused by our
definition of minimum cell luminosity to constitute X-ray bright cells. 
There are almost no high-temperature clusters with $T_x \ga 10{\rm keV}$.
Again, this is due to the limited volume of our computational domain
which is not big enough to contain high-temperature clusters.
The observed temperature distribution function is $(1.8_{-1.5}^{+0.8} \times
10^{-3} h^3 {\rm Mpc}^{-3} {\rm keV}^{-1})(kT)^{-4.7 \pm 0.5}$ according to 
Henry \& Arnaud (1991), and shown as the boxed area.
In the reliable range of $T_x$, the distribution functions for the both
models are almost identical.
This is expected, since $T_x$ is mainly determined by the gravitational
potential of {\it total} matter.
Both the distribution functions agree well with the observed data.

The relation between the cluster's temperature, $T_x$, and luminosity,
$L_x$, should be less sensitive to the normalization of power spectrum.
In Figure 6, we show the plots of $T_x$ versus $L_x$ at $z = 0$.
The upper panel is for CHDMa and the lower panel is for CHDMb.
The boxed area indicates the observed relation, $\log_{10}T_x({\rm keV})
= \log_{10}(4.2_{-0.8}^{+1.0}) + (0.265 \pm 0.035) \log_{10}(h^2 L_{44})$
according to Henry \& Arnaud (1991).
The best straight-line fits, $\log_{10} kT_x = A + B \log_{10} L_x$,
are shown as a dotted line for CHDMa and solid line for CHDMb.
$A = 0.672$ and $B = 0.413$ for CHDMa, and $A = 0.495$ and $B = 0.466$
for CHDMb, respectively.
For a given luminosity, the temperature is somewhat lower for CHDMb
($A = 0.495$) than for CHDMa ($A = 0.672$), and the slopes ($B = 0.413$
and $0.466$) are somewhat steeper than that indicated by observations
($B = 0.265 \pm 0.035$).
However, the observed slope is determined  mainly by high-luminosity clusters.
We find that in the region where comparison can be made, 
the agreement with the observations is better for CHDMa than CHDMb.

\begin{figure}
\caption{Relation of $T_x$ versus $L_x$ at $z = 0$.
The open circles in the upper panel are from CHDMa, and the filled circles
in the lower panel is from CHDMb.
The dotted and solid lines show the best fits, and the boxed area indicates
the observational data (Henry \& Arnaud 1991).}
\end{figure}

For each simulated X-ray cluster, the gas mass and the total mass within
a sphere of radius $1 h^{-1} {\rm Mpc}$ are calculated at $z = 0$.
The histogram of the ratio of the two masses is shown in Figure 7.
The thick dotted histogram shows the ratio distribution for CHDMa, and
the thick solid shows the ratio distribution for CHDMb.
These histograms are arbitrarily normalized to have a peak height similar
to the observation data, which is drawn with the thin solid histogram,
adopted from Jones \& Forman (1992).
The more recent observed ratio of the gas to total mass is 
$(M_{gas}/M_{tot})_{obs} = 0.1 - 0.22$ for a refined sample of
$13$ clusters (White \& Fabian 1995), so the observed ratio has 
gone up a little.
The median of the computed ratio is $(M_{gas}/M_{tot})_{com} =
0.0731 \pm 0.0038$ for CHDMa and $(M_{gas}/M_{tot})_{com} =
0.120 \pm 0.0069$ for CHDMb. 
Improving observations will probably narrow the observed histogram, while
increasing the dynamical range of numerical simulations and incorporating
more realistic physics will widen the computed histograms.
Both expected improvements should make the agreement better.

\begin{figure}
\vspace{0.75truein}
\epsfysize=2.4in\epsfbox[160 300 460 500]{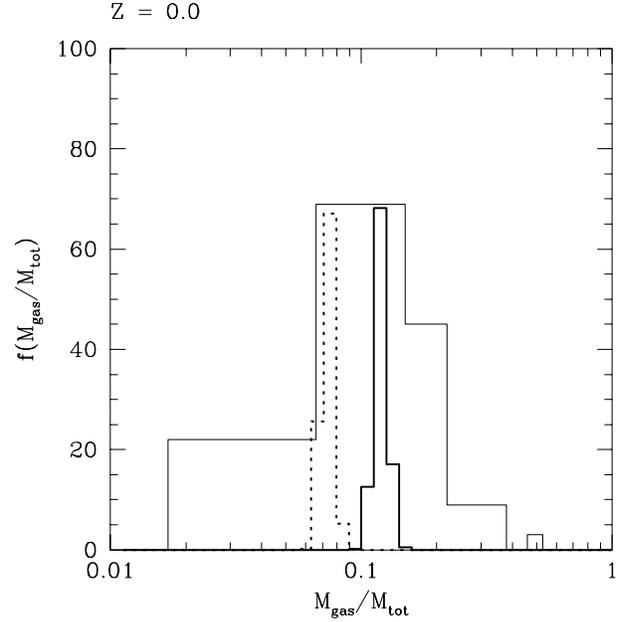}
\vspace{0.1truein}
\caption{Histogram of the ratio of gas to total mass in the X-ray
clusters with $L_x > 10^{41} {\rm erg}~{\rm s}^{-1}$ at $z = 0$.
The thick dotted line represents the histogram from CHDMa, and the thick
solid shows the histogram from CHDMb.
The thin solid line shows the observational data (Jones \& Forman 1992).}
\end{figure}

Figure 8 shows the evolution of the average X-ray temperatures
for clusters with $L_x > 10^{43} {\rm erg}~{\rm s}^{-1}$ from $z = 0$ 
to $z = 1.5$.
Open circles are for CHDMa and filled circles are for CHDMb. 
Also shown are the best-fits to the evolution analytically predicted by
Kaiser (1986), $T_x \propto (1 + z)^{-1}$ for $\Omega = 1$.
The evolutions for the both model are similar, as expected.
But there is an overall difference that the average temperatures are
smaller in CHDMb by a factor of $0.65$ than those in CHDMa. 
The computed evolution of the average temperatures agrees only
marginally with analytical prediction by Kaiser (1986).

\begin{figure}
\vspace{0.7truein}
\epsfysize=4.7in\epsfbox[160 200 460 600]{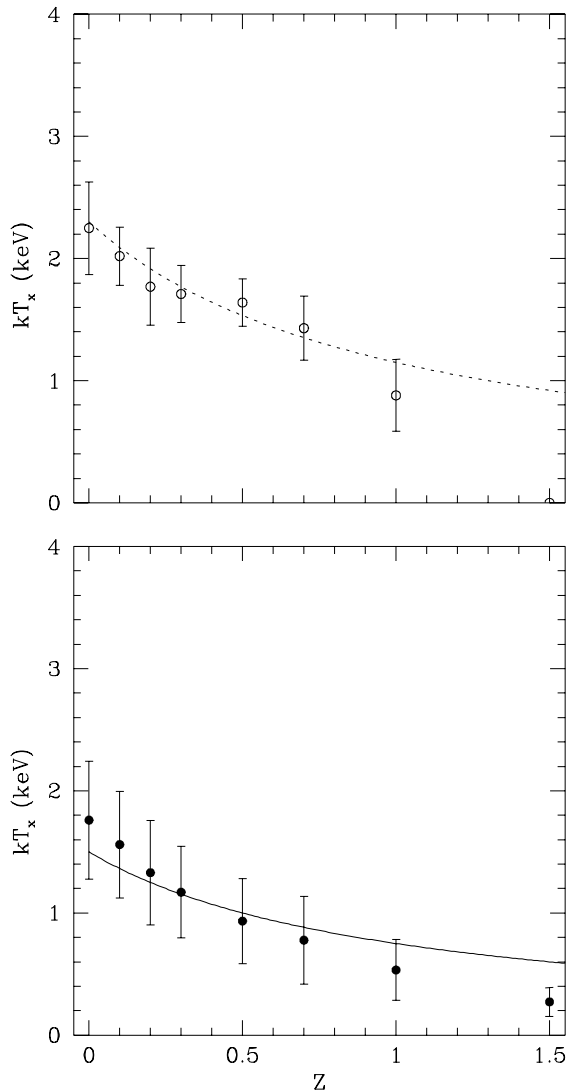}
\vspace{0.3truein}
\caption{Average X-ray temperature for the clusters with
$L_x > 10^{43} {\rm erg}~{\rm s}^{-1}$ as a function of redshift.
The open circles in the upper panel is from CHDMa, and the filled circles
in the lower is from CHDMb.
The dotted and solid curves show the best-fits to the evolution predicted 
by Kaiser (1986), $T_x\propto(1+z)^{-1}$.}
\end{figure}

\section{Conclusion}

We have computed the linear power spectrum of total matter for
spatially flat, {\it COBE}-normalized CHDM models.
We have found that the models with $h/n/\Omega_h=0.5/0.9/0.1$ and
$0.5/0.9/0.2$ give a power spectrum in good agreement with the
observed power spectrum.
Then, through numerical simulations that include the hydrodynamics of
baryonic matter, as well as the particle dynamics of dark matter,
we have computed the properties of X-ray clusters for the models with
$h/n/\Omega_h=0.5/0.9/0.2$ and $\Omega_b=0.05$ and $0.1$.
We have found that the models with $\Omega_b = 0.05$ produce X-ray
clusters with properties that agree well with the observed data.

Our results have shown that the physical properties of X-ray clusters
considered in this paper are quite sensitive to $\Omega_b$, since the
bremsstrahlung emission is proportional to $\rho_b^2$.
Although the number of X-ray bright clusters in the model with
$\Omega_b = 0.1$ exceeds slightly the observed number, we still consider
the model to be marginally consistent with observations.
However, the {\it COBE}-normalized CHDM model with $\Omega_b>0.1$ may
be ruled out by the present work.
Finally, we note that that our simulations are not fully converged, so
X-ray luminosity may have been under-estimated by a factor of $3$ or more.
Hence, future simulations with higher resolution will put a more stringent
constraint on $\Omega_b$.

\section*{acknowledgments}

This work is based on the Master's thesis of EC at Kyungpook National
University in Korea.
The simulations were performed on Cray-C90 at Systems Engineering
Research Institute in Korea.
We are grateful to Drs. H. Kang and J. Hwang for discussions, and the
referee, Dr. J. Primack, for suggestions and comments on the manuscript.
The work by EC was supported in part by KOSEF grant No. 961-0203-013-1.
The work by DR was supported in part by Seoam Scholarship Foundation
and by KOSEF through the 1997 Korea-US Cooperative Science Program.

\bsp

\end{document}